\documentclass[preprint,12pt,authoryear]{elsarticle}
\usepackage{mathptmx} % times new roman
\ifx\pdfoutput\undefined
\usepackage[dvips]{graphicx}
\else
\usepackage{graphicx}
\fi
\usepackage{epstopdf}
\usepackage{floatrow}
\usepackage{setspace} % double spacing

\newfloatcommand{capbtabbox}{table}[][\FBwidth]

\usepackage{multirow} 
\usepackage{dsfont}
\usepackage{amsmath}
\usepackage{amssymb}
\usepackage{amsthm}
\usepackage{mathtools}
\usepackage{amsfonts}
\usepackage{natbib}
\usepackage{caption}
\usepackage{multirow}
\usepackage{soul}
\usepackage[T1]{fontenc}
\usepackage{blindtext}
\usepackage{epsfig}
\usepackage{bm}
\usepackage{booktabs}
\usepackage{float}
\usepackage{algorithmic}

\usepackage{float}
\usepackage{subcaption}
\usepackage[ruled,vlined]{algorithm2e}
\floatstyle{plaintop}
\restylefloat{table}
\usepackage{graphicx}
\usepackage[utf8]{inputenc}
\usepackage[english]{babel}
\usepackage[normalem]{ulem}

\usepackage[font=small,labelfont=bf]{caption}\usepackage[pdftex,dvipsnames,usenames]{color}% % % % % % % %  
\usepackage{sectsty}
\allsectionsfont{\usefont{OT1}{phv}{bc}{n}\selectfont} 

\long\def\symbolfootnote[#1]#2{\begingroup%
	\def\thefootnote{\fnsymbol{footnote}}\footnote[#1]{#2}\endgroup}

\begin{document}
\doublespacing %double spacing
\begin{frontmatter}

\title{\textbf{Determinants of financial and digital inclusion in West and Central Africa: Evidence from binary models with cross-validation}}
\author[1]{ Ismaila A. Jallow \corref{cor1}}
\author[1,*]{Samya Tajmouati
\corref{cor1}}
%\ead{samya.tajmouati@gmail.com}

\address[*]{Corresponding author: Samya Tajmouati, s.tajmouati@um5r.ac.ma}

    %``````````````````````````````````````````````````````%
	% Abstract                                             %
	%``````````````````````````````````````````````````````%
\begin{abstract}

This study examines the determinants of financial and digital inclusion in West and Central Africa using the World Bank Findex 2021 data. Unlike prior works that rely solely on traditional logit and probit models, we combine country-by-country analysis with robustness checks including K-fold cross-validation and Vuong test. Three samples were considered  : a full sample combining both regions and two separate subsamples for West and Central Africa. The results indicate that gender, educational attainment, income level, and place of residence are significant factors influencing both financial and digital inclusion in the full sample and the West African subsample. In the Central African subsample, gender is not significant; however, age, education, income, and rural residence emerge as key determinants of financial and digital inclusion. Overall, Ghana stands out as the most financially inclusive country, although it trails Senegal in terms of credit access and digital payment use. Nigeria leads in formal account ownership and formal savings but falls considerably behind Ghana in mobile money account ownership and digital payments. Central African countries generally report lower levels of inclusion compared to West Africa, with Cameroon performing comparatively better. 

\end{abstract}

\end{frontmatter}
\makeatletter
\def\ps@pprintTitle{%
  \let\@oddhead\@empty
  \let\@evenhead\@empty
  \def\@oddfoot{\reset@font\hfil\thepage\hfil}
  \let\@evenfoot\@oddfoot
}
\makeatother
%~~~~~~~~~~~~~~~~~~~~~~~~~~~~~~~~~~~~~~~~~~~~~~~~~~~~~~~~~~
%% Introduction

\section{Introduction}
An Africa where every person has access to financial services, whether through owning a formal account, saving and borrowing via formal channels, or having the education needed to participate in digital payments, has been a key goal for many African nations. This vision aligns with the G20 2010 summit agenda, which stresses promoting financial inclusion across the continent \citep{demirgucc2017financial}.  Financial inclusion is an important driver of sustainable development \citep{ozili2022financial}. As it reduces inequality and poverty levels and encourages efficient allocation of capital funds by mobilizing savings and managing risk \citep{rashdan2020determinants}. In the classical Keynesian model of Harrod-Domar model, the main strategy for economic development is savings mobilization and generation of investment \citep{domar1947expansion,harrod1939essay}. Hence, they affirm that savings encourage investments and which leads to economic growth. According to \cite{demirgucc2022global}, financial inclusion is defined as the access and the usage of regulated financial services including savings, credit, insurance, and payments, particularly targeting marginalized groups like women, rural populations and low-income households. Digital inclusion, on the other hand, refers to the adoption of technology-driven financial services such as mobile money, digital wallets, and fintech innovations that triumph over traditional barriers through the expansion of digital infrastructure and innovations. Looking at the past decades, the overall financial and digital inclusion has come a long way, as \citep{demirgucc2012financial,demirgucc2012measuring} provide one of the first foundational global insights using cross-sectional logistic and linear regression models on the 2011 Global Findex data. They reported that approximately 50\% of adults aged 15 and older globally have a formal financial account, compared with only 41\% in developing countries and just 23\% in Africa. They added that limited funds in which 32\% of unbanked adults cited as the main cause of their financial exclusion, followed by high costs at 25\%, long distances at 30\%, insufficient documentation at 22\%, and mistrust at 18\% are the top barriers preventing them from accessing financial services. A decade later \citep{demirgucc2022global} reveals significant progress, as the global account ownership increased to 62\% and 53\% in developing countries, sub-Saharan Africa registered a 55\% increment, that a huge boost considering it was just at 23\% in the 2011’s.  Mobile money emerged as the catalyst to this revolution, as 33\% of adults aged 15 years and older in the regions are using these services by 2021, compared to just 12\% in 2014. \cite{allen2014african} use case-control and econometric analysis of M-Pesa adoption in Kenya to demonstrate how mobile technology transformed financial access; their results show an increasing account ownership from 27\% to 56\% among low-income households between 2009-2013. \cite{anson2013financial} also showed that post offices are more likely than traditional financial institutions to provide accounts to groups such as the poorest, the less educated, and the disabled. Despite the overall formal account ownership improvements, a large number of adults remain excluded from formal financial services, the reason could be voluntary or involuntary \citep{carpenter2002household,ezzahid2022financial,lee2016financing}. About 98 million sub-Saharan adults participate in (Rotating Savings and Credit Associations, ROSCAs), 58\% being women, and 38\% borrow from friends or family in a year, which indicates a strong reliance on informal practices, particularly among women and rural populations \citep{klapper2013financial}.\\ 
The aim of this article is to identify the determinants of financial and digital inclusion in West and Central Africa. More specifically, the study seeks to determine the overall factors influencing financial and digital inclusion across both regions combined and then proceed with a separate analysis for each sub-region. The objective is to conduct three distinct sample analyses and carry out a comparative assessment between them. It’s important to note that factors determining financial inclusion also mirror the factors determining digital inclusion \citep{demirgucc2022global}.\\
Empirically, a lot of studies carried out in different countries around the world have examined and analyzed various factors determining financial and digital inclusion.  \cite{allen2016foundations} used the probit models on the 2011 World Bank Global Findex data. Their results reveal that older, employed, married, wealthier, better-educated, and urban individuals are more likely to own formal financial accounts, save through official channels, and borrow from formal institutions. On the same Findex database in a case in China, \cite{fungavcova2015understanding} used a probit regression. They also found that older, wealthier, and more educated men are more financially included. Those in the lowest income quintile cited lack of funds and another member of their family having an account as barriers, while the more educated and those in the highest income quintile are more concerned about costs and system reliability. \cite{aterido2013access}, \cite{demirgucc2012financial}, and \cite{demirgucc2012measuring} utilize multivariate logit and probit regression across multiple countries to find the reason for gender-based exclusion. They identify cultural norms, documentation issues, and household dynamics as the main reasons that stop women from using financial services. They added that an individual's income and education influence a lot when it comes to accessing both formal and informal credit. Women have been most of the time more often affected by low income and education; they tend to go for informal credit more compared to formal ones. In expansion on the gender-related issues in financial inclusion, \cite{aterido2013access} analyzed nine sub-Saharan African countries and ended up finding no direct gender discrimination. They found that women's lower access to formal services, like owning a formal account, saving or borrowing formally, is due to the high exclusion they face when it comes to access to education and employment. For those reasons, women tend to lend more to informal financial services. The demographic factors were analyzed by \cite{soumare2016analysis}, where they used the 2011 findex database, and  applied cluster-specific fixed effects on both probit and logit models in West and Central Africa. As usual, their results indicate that access to formal finances in both regions depends on individual characteristics such as gender, education, age, income, etc. However, they confirm some disparities are unique to a specific region, as they found that in Central Africa, being male, married, educated, and living in an urban area, are more likely to hold formal accounts, and in West Africa, income and household size matter more. Looking at Africa as a whole, \cite{zins2016determinants} used a probit regression model on the 2014 Findex database, and they also found that wealth, education, age, and gender significantly affect financial inclusion, with poorer, less educated women facing the greatest barriers, \cite{tuesta2015financial} also identify the same determinants in Argentina through a national household survey. \cite{akudugu2013determinants} used a logit model on Ghanaian survey data they identify literacy, lack of funds, documentation, and distance as significant barriers to financial inclusion. This finding is echoed by \cite{sarma2011financial}, who use panel regression across 49 developing countries to link financial inclusion positively with human development indicators. \cite{suri2016long} use difference-in-differences analysis on Kenyan account data to show that M-Pesa caused a statistically significant increase in consumption and poverty reduction, especially among female-headed households, demonstrating mobile money’s tangible economic benefits. \\
While all these studies provide an extensive overview of financial and digital inclusion around the world, very few have focused on the West and Central African regions. They are often cited as the least financially inclusive in sub-Saharan Africa. Despite this setback, they host some of the continent’s most advanced monetary and customs unions \citep{soumare2016analysis}. Although some literature offers a broad comparison between the two regions, few studies analyze financial inclusivity on a country-by-country basis. Traditional models such as probit and logit are widely used in the literature due to the binary nature of most financial inclusion indicators. However, to our knowledge,  no existing study combined the most recent Global Findex 2021 dataset - which captures financial and digital dynamics during and after after COVID-19 -  with  machine learning techniques such as K-fold cross-validation to assess variable overfitting \citep{refaeilzadeh2009cross,tajmouati2024applying}.  \\
Our paper contributes to the literature in three ways. First, we use the latest Findex 2021 data  including 17 countries in West and Central Africa and 8 dependent variables of financial and digital inclusion \citep{demirgucc2022global}. Second, we strengthen the methodology by applying K-fold cross-validation and Vuong test for model selection \citep{vuong1989likelihood}. Third, we move beyond regional averages by estimating models over three separate samples (West and Central Africa, West Africa, and Central Africa) and conducting country-by-country comparisons.This allows us to extract meaningful determinants of financial and digital inclusion and propose strong policy recommendations for individual countries and sub-regions.
% To contribute to the existing literature, we employ the latest Findex 2021 data , using a sample of 17 countries from West and Central Africa and eight dependent variables ranging from formal account ownership to formal and informal savings and borrowings, as well as inclusive digital payments. As part of our robustness checks, we apply K-fold cross-validation \citep{refaeilzadeh2009cross,tajmouati2024applying}. Guided by the Vuong test \citep{vuong1989likelihood}, we perform three separate logit and probit models: (i) a full-sample model covering both West and Central African countries, (ii) a sub-sample model for only West African countries, and (iii) a sub-sample model for only Central African countries. Additionally, we analyze the current determinants of digital inclusion, and our binary models go beyond merely identifying socio-economic and demographic factors by comparing individual countries to a chosen reference country. This approach enables us to extract meaningful insights into the inclusiveness of each country and to propose more targeted policy recommendations across the sub-regions.\\
Our results show that in both the full sample and the West African subsample, the key significant determinants of financial and digital inclusion include gender, age, residence, education, and income level, except for formal borrowing and informal savings, which do not show gender-related significance. We also found that age is not significant when it comes to digital merchant payments in both the full and the West African subsamples. On the other hand, in the Central African subsample, gender is not statistically significant in determining financial and digital inclusion. This is due to the deeper socio-economic inequalities and widespread exclusion linked to limited access to education, employment, and low income for households in the region. However, age, employment status, residence, education, and income remain significant in Central Africa, while formal borrowing shows mixed significance across the independent variables. Using Ghana as the reference country in the full sample, it emerges as the most financially inclusive, with strong adoption of both financial and mobile money accounts. However, Ghana lags behind several other countries when it comes to both formal and informal access to credit and also trails behind Senegal in merchant digital payments. In the West African subsample, where we used Senegal as the reference, Ghana outperforms Senegal on most indicators except credit access and digital merchant payments. Nigeria demonstrates the strongest performance in formal account ownership and financial savings compared to all other countries in this study, but falls behind both Ghana and Senegal in mobile money and digital payment adoption. When we compared the two sub-regions, Central African nations are generally less financially inclusive in terms of digital payments, formal savings and borrowing, and account ownership. Though there are still many gaps, Cameroon stands out as a relative leader in both financial and digital inclusion in a modest way. Our study also shows that access to credit behaves differently from other financial inclusion-dependent indicators. Interestingly, Ghana and Senegal, which show the highest levels of both financial and digital inclusion, lag behind countries like Liberia, Chad, Cameroon, Mali, etc., when it comes to both formal and informal credit accessibility, which may explain the low sensitivity observed in the K-fold cross-validation test. 
%Lastly, ownership of a financial account does not necessarily translate into higher digital payments in the two regions, but the adoption of mobile money does.\\
The rest of this paper is structured as follows. Section~\ref{sec.2} presents the data and variables. Section~\ref{sec.3} presents the methodology adopted. Section~\ref{sec.4} presents the results and discussion. 
Finally, Section~\ref{sec.5} presents the conclusions as well as the policy implications of the paper.

%%% For Now I have corrected spelling and grammar error in the abstract and the introduction but still need to re-write them.
%~~~~~~~~~~~~~~~~~~~~~~~~~~~~~~~~~~~~~~~~~~~~~~~~~~~~~~~~~~
%% Conformal prediction

\section{Data and variables}
\label{sec.2}
\subsection{\textit{Data source}}
We use the Global Findex Database 2021, based on surveys of 128,000 adults in 123 economies during the COVID-19 pandemic. It provides updated insights into financial inclusion, digital payments, and financial resilience. Our sample focuses on 17 countries across western and central Africa. The sample consists of 4 central African countries: Chad (TCD), Democratic Republic of Congo (COD), Cameroon (CMR) and the Republic of Congo (COG) out of the 5 included in the 2021 Findex Database, and 13 west African countries Senegal (SEN), The Gambia (GMB), Guinea (GIN), Ghana (GHA), Mali (MLI), Burkina Faso (BFA), Serra Leon (SLE), Liberia (LBR), Togo (TGO), Benin (BEN), Niger (NER), Nigeria (NGA) and Ivory Coast (CIV). Original observation for our 17 country sample consists of 17000 observations with 1000 observations per country, but due to data cleaning in which missing data and unused variables are excluded, our final data sample ended up being 16,736 observations with 1.6\% data reduction from the original observation.
\subsection{\textit{Dependent Variables}}
\begin{itemize}
    \item \textbf{Financial account ownership $(account\_fin)$: }Takes 1 if a person i in country j had an account at a bank or another type of financial institution, like a credit card, credit union, microfinance institution, cooperative, or post office, and 0 otherwise.
    \item \textbf{Mobile money account ownership $(account\_mob)$:} Returns 1 if a person i in country j has a mobile money account, and 0 otherwise.
    \item \textbf{Saved using an account at a financial institution $(saved\_fin)$: } As part of our data cleaning process, we only considered responses coded as 1 and 2, which were then transformed into a binary (dummy) variable: 1 and 0, respectively. Responses coded as 3 and 4, which indicate “don’t know” and “refuse to answer,” were excluded. Based on this new dummy variable, an individual i in country j takes the value 1 if they reported having saved or set aside money in the past 12 months using an account at a bank or other formal financial institution (excluding mobile money), and 0 otherwise.
    \item  \textbf{Inclusive saving behaviors $(saved)$: } Accounts for both the formal and the informal sectors; it takes 1 if a person i in country j  actually saved or set aside money in the previous 12 months, whether through a financial institution account, a mobile money account, a savings club, a non-family member, or for any other reason, and 0 otherwise. 
    \item \textbf{Borrowed from a financial institution $(borrowed\_fin)$: } This variable also underwent the same data cleaning process as $saved\_fin$. An individual i in country j takes the value 1 if they reported having borrowed money from a bank or other formal financial institution (excluding mobile money) in the past 12 months, and 0 otherwise.
    \item \textbf{Inclusive borrowed behaviors $(borrowed)$: } Accounts for both the formal and the informal sectors; it takes 1 if a person  i in the country j borrowed money in the previous 12 months, either alone or in conjunction with another person, from a bank or other financial institution through a mobile money account, from a friend or family member, from an unofficial savings group, or for any other reason, and 0 otherwise.
    \item \textbf{Made a digital merchant payment $(merchantpay\_dig)$: } Takes 1 if respondent used a debit or credit card or a mobile phone to pay for an online purchase or make a purchase in-store during the previous 12 months, and 0 otherwise.
    \item  \textbf{Inclusive digital payments $(anydigpayment)$: } Takes 1 if, during the previous 12 months, a person i from country j  used digital payment methods like mobile money, debit or credit, or internet payments for such transactions as bill payment, online or in-store purchases, remittances, government transfers, wages, or pensions, and 0 otherwise.
    
\end{itemize}

\subsection{\textit{Independent variables}}
\begin{itemize}
    \item \textbf{Female $(female)$: } Takes 1 if the respondent is a female, and 0 otherwise. We expect it will be more difficult for a woman than a man to own an account and have access to financial services.
    \item \textbf{Age $(age)$: } Refers to the age of the individual; we expect this to be in line with past findings that young and old people face higher financial exclusion.
    \item \textbf{Age$^2$ $(age.2)$} : This is age squared; it captures the non-linear effect associated with age. 
    \item \textbf{Rural $(rural):$} A dummy that takes the value of 1 if the respondent lives in a rural area and 0 otherwise. The literature suggests that access to financial services is more challenging for individuals living in rural areas in Africa.
    \item \textbf{Unemployed $(emplout)$} : Takes 1 if the respondent is out of the workforce and 0 otherwise.
    \item \textbf{Education $(educ) :$} This variable captures the three levels of education used in this study: primary or less $(educ\_1)$, secondary $(educ\_2)$, and tertiary or higher $(educ\_3$).  We expect a positive correlation between higher levels of education and the likelihood of using financial services. Due to the categorical nature of this variable, we applied one-hot encoding. To address the issue of perfect correlation and avoid multicollinearity, we included only $educ\_1$ and $educ\_2$ in the regression, using $educ\_3$ as the reference group against which all other education levels are compared. 
    \item \textbf{Income $(inc) :$} Income level is represented using quintiles, dividing the population into five equal groups. These include the poorest 20\% ($inc\_1)$, the second quintile $(inc\_2)$, the third quintile $(inc\_3)$, the fourth quintile $(inc\_4)$, and the richest 20\% $(inc\_5)$. Given the categorical nature of this variable, it was treated similarly to the education variable. To prevent perfect multicollinearity, we included $inc\_1 $ through $inc\_4$ in the regression, with $inc\_5$ serving as the reference group. We expect a positive correlation between income level and financial inclusion. 
    \item \textbf{Country $(economycode)$ : } This variable identifies the country of residence for each individual in the sample. As it is categorical, it was treated similarly to the education and income variables. One country is always designated as the reference category to avoid perfect multicollinearity before running regressions. The variable includes 17 categories if the full sample is considered, 4 categories if only the Central African countries are considered and 13 categories if only the West African countries are considered. For each category, the variable takes the value 1 if the individual resides in that country, and 0 otherwise.
\end{itemize}

\subsection{\textit{Descriptive statistics }}
\begin{figure}
    \centering
    \includegraphics[width=1\linewidth]{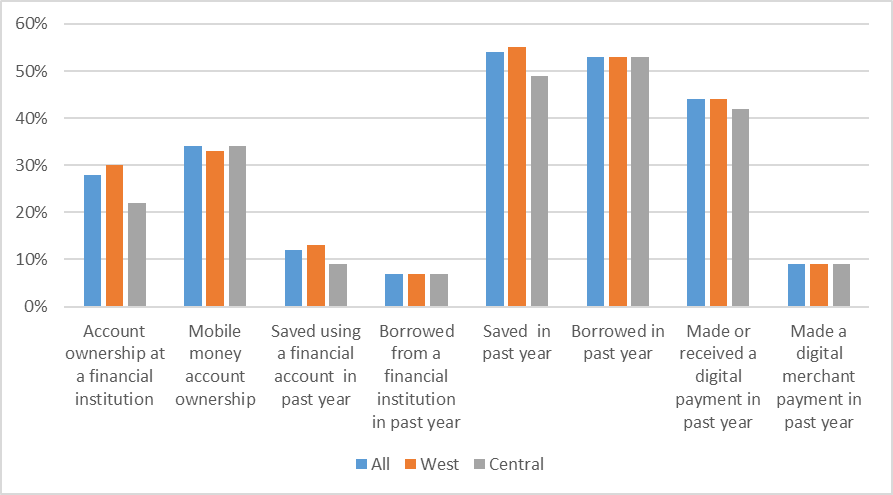}
    \caption{ Distribution of financial and digital inclusion indicators in Central and West Africa}
    \label{fig1}
\end{figure}

\begin{figure}
    \centering
    \includegraphics[width=1\linewidth]{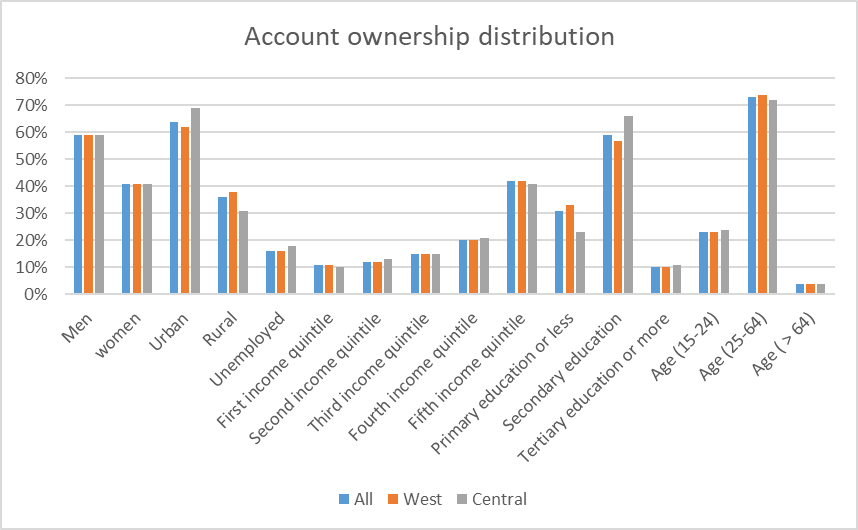}
    \caption{Account ownership distribution by sociodemographic characteristics in West and Central Africa}
    \label{fig2}
\end{figure}

\begin{figure}
    \centering
    \includegraphics[width=1\linewidth]{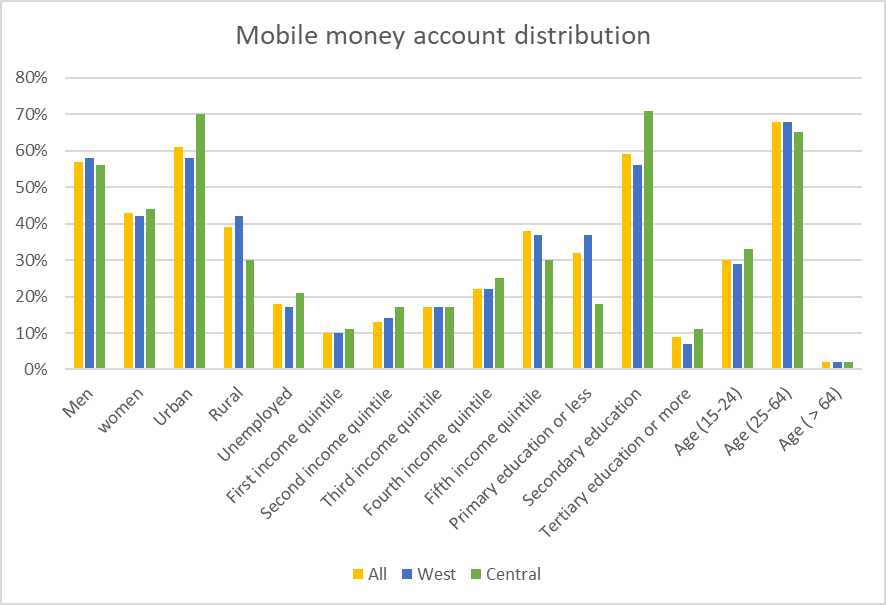}
    \caption{Mobile Money Account ownership distribution by sociodemographic characteristics in West and Central Africa}
    \label{fig3}
\end{figure}

\begin{figure}
    \centering
    \includegraphics[width=1\linewidth]{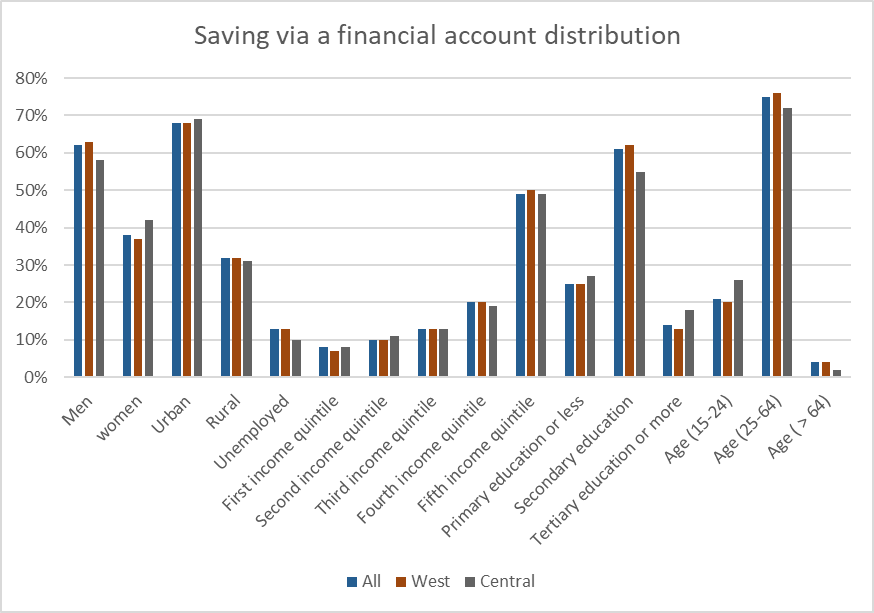}
    \caption{Financial savings distribution by sociodemographic characteristics in West and Central Africa}
    \label{fig4}
\end{figure}

\begin{figure}
    \centering
    \includegraphics[width=1\linewidth]{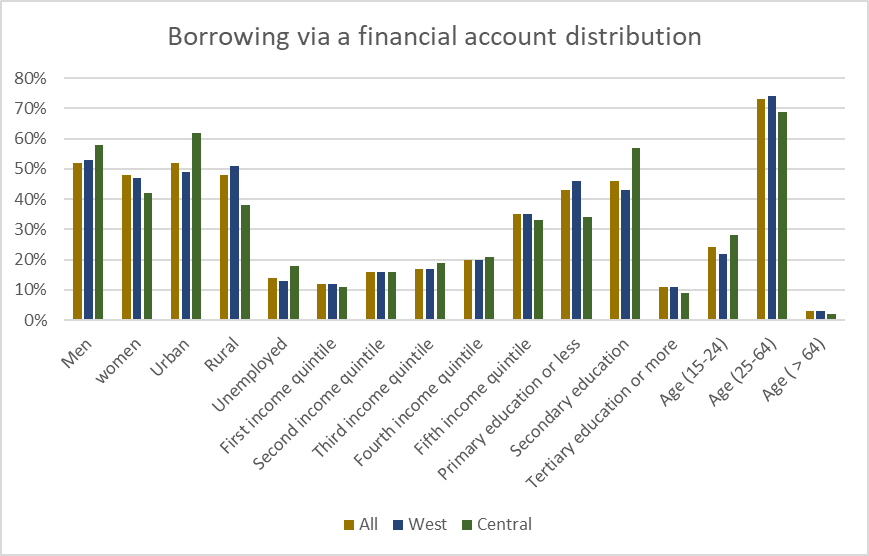}
    \caption{Financial borrowings distribution by sociodemographic characteristics in West and Central African}
    \label{fig5}
\end{figure}

\begin{figure}
    \centering
    \includegraphics[width=1\linewidth]{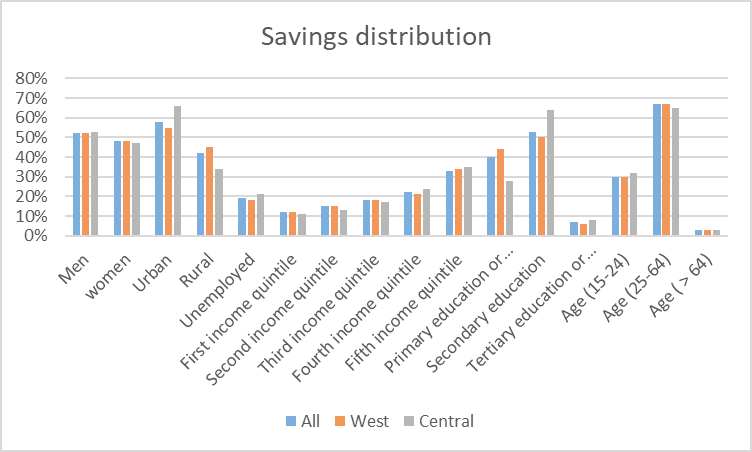}
    \caption{Savings distribution by sociodemographic characteristics in West and Central Africa}
    \label{fig6}
\end{figure}

\begin{figure}
    \centering
    \includegraphics[width=1\linewidth]{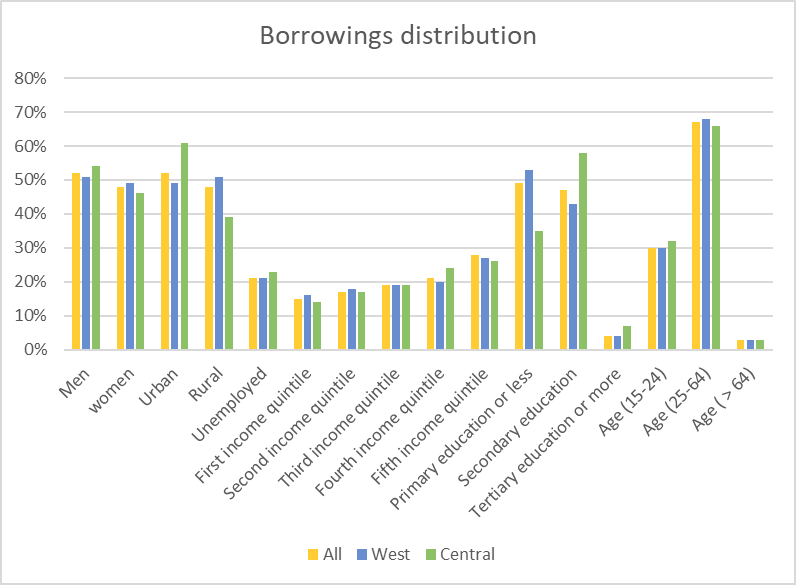}
    \caption{Borrowings distribution by sociodemographic characteristics in West and Central Africa}
    \label{fig7}
\end{figure}

\begin{figure}
    \centering
    \includegraphics[width=1\linewidth]{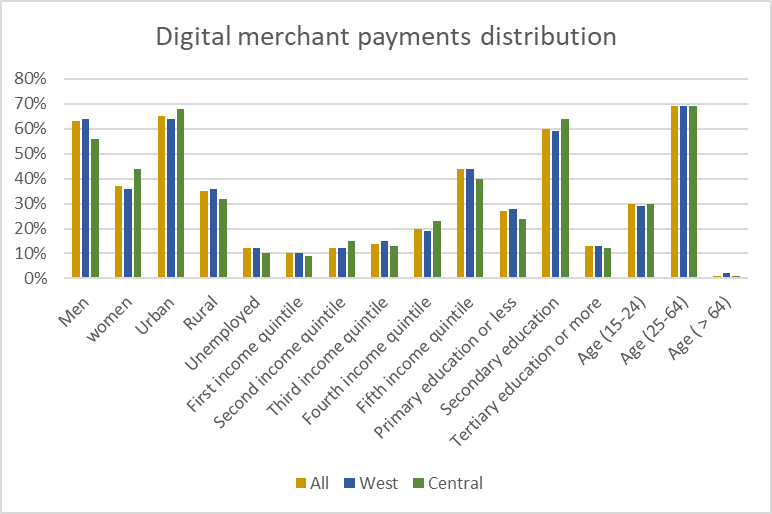}
    \caption{Digital merchant payments distribution by sociodemographic characteristics in West and Central Africa}
    \label{fig8}
\end{figure}

\begin{figure}
    \centering
    \includegraphics[width=1\linewidth]{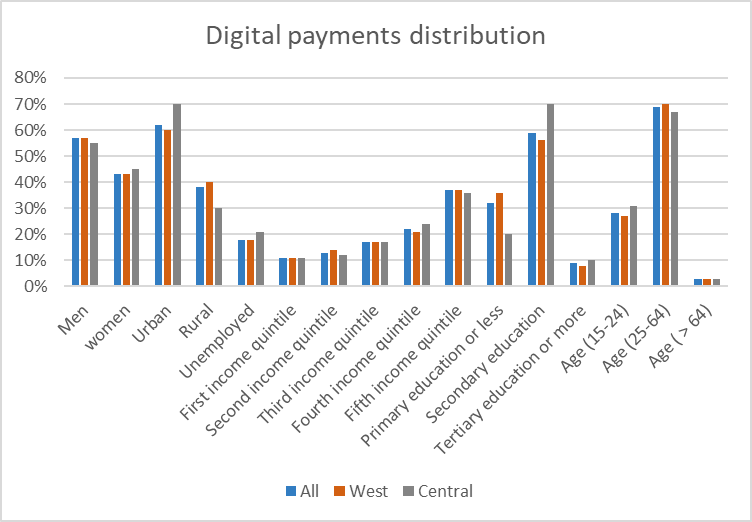}
    \caption{Digital payments distribution by sociodemographic characteristics in West and Central Africa}
    \label{fig9}
\end{figure}

Figure~\ref{fig1} shows that about 28\% of respondents in West and Central Africa have a financial account, with around 30\% in West Africa and 22\% in Central Africa. Saving through a financial institution is less common in Central Africa, with approximately 9\%. On the other hand, saving in the past year, whether formal or informal, is more common, reaching 54\% in both regions. Borrowing from a financial institution remains limited in both regions, at approximately 7\%. In contrast, borrowing in the past year, whether formal or informal, is widespread, reaching 53\% in both regions.  Digital and merchant digital payments show similar levels in both regions, reaching 44\% and 9\% respectively. \\
Figures~\ref{fig2},~\ref{fig3},~\ref{fig4}~\ref{fig5},~\ref{fig6},~\ref{fig7},~\ref{fig8}, and ~\ref{fig9} provide an overview of financial behaviors across sociodemographic characteristics in West and Central Africa. From figures~\ref{fig2} and~\ref{fig3}, women, rural, unemployed, and poorest quintile report the lowest financial account ownership, mobile money account ownership, digital and merchant digital payments, and saving behaviors in West and Central Africa.  The sociodemographic levels are almost the same for borrowing variables, with the exception of age, where individuals up 64 consistently show lower rates across all financial and digital indicators.


\section{Methodology}
\label{sec.3}

To account for the regional diversity in the determinants of financial inclusion and digital payments, we run separate binary regression models based on different geographical scopes; three separate samples are considered :
\begin{itemize}

    \item \textbf{Full Sample  : } : It includes all countries in the dataset, as detailed in the data section. Ghana is selected as the reference country to allow for meaningful comparisons across nations. This approach enables the identification of general patterns and disparities in financial inclusion and digital payments across the entire region.

    \item  \textbf{West African Subsample:  }It focuses exclusively on West African countries. Senegal serves as the reference country, facilitating a more focused analysis of regional dynamics and inter-country differences within West Africa.
   \item \textbf{Central African Subsample:  }In this case, only Central African countries are included. Cameroon is used as the reference country to explore the specific drivers and barriers to financial inclusion in Central Africa and to highlight intra-regional disparities.
    
\end{itemize}

For each sample, we begin by determining the most appropriate binary model, logit or probit, for each dependent variable using the Vuong test. Once the models are selected, we assess their robustness using K-fold cross-validation. 

\subsection{\textit{Vuong test}}
We apply the Vuong test \citep{vuong1989likelihood} to identify the best model –probit or logit - for each dependent variable. Under the null hypothesis of model equivalence, if the test statistic exceeds 1.96, the probit model is selected. Otherwise, if the test statistic is below  -1.96, the logit model is chosen. Otherwise, we cannot distinguish between models.
\begin{table}[h]
    \centering
    \begin{tabular}{c|c|c|c}
        Dependent variable & z-statistic & p-value & Preferred model \\ \hline
        $account\_fin$ & -4.889     & 0.0000 & Logit \\
        $account\_mob$ & -2.055    & 0.0199 & Logit \\
        $saved\_fin$ & -2.661    & 0.0039 & Logit \\
        $borrowed\_fin$ & 2.540    & 0.0055 & Probit \\
        $saved$ & -0.170   & 0.4326 & Equal \\
         $borrowed$ & -2.492    & 0.0064 & Logit \\
          $merchantpay\_dig$ & -0.246    & 0.403 & Equal \\
          $anydigpayment$ & 1.506    & 0.0661 & Equal \\ \hline
    \end{tabular}
    \caption{Vuong test results on dependent variables – Full sample }
    \label{tab1}
\end{table}
\begin{table}[h]
    \centering
    \begin{tabular}{c|c|c|c}
        Dependent variable & z-statistic & p-value & Preferred model \\\hline
        $account\_fin$ & -3.911     & 0.0000 & Logit \\
        $account\_mob$ & -1.437    &0.0753 & Equal \\
        $saved\_fin$ & -2.984    & 0.0014 & Logit \\
        $borrowed\_fin$ & 0.706    & 0.7600 & Equal \\
        $saved$ & 3.999    & 0.0000 & Probit \\
         $borrowed$ & -2.095    & 0.0181 & Logit \\
          $merchantpay\_dig$ & 1.018    & 0.8456 & Equal \\
          $anydigpayment$ & -0.236    & 0.4068 & Equal \\\hline
    \end{tabular}
    \caption{Vuong test results on dependent variables – West African Subsample }
    \label{tab2}
\end{table}
\begin{table}[h]
    \centering
    \begin{tabular}{c|c|c|c}
        Dependent variable & z-statistic & p-value & Preferred model \\ \hline
        $account\_fin$ & -2.270     & 0.0116 & Logit \\
        $account\_mob$ & -1.710    &0.0436 & Logit \\
        $saved\_fin$ & 0.461    & 0.6776 & Equal \\
        $borrowed\_fin$ & 0.290   & 0.6139 & Equal \\
        $saved$ & -1.520    & 0.0642 & Equal \\
         $borrowed$ & -1.388    & 0.0825 & Equal  \\
          $merchantpay\_dig$ &2.415& 0.9921 & Probit \\
          $anydigpayment$ & 0.326    & 0.6277& Equal \\ \hline
    \end{tabular}
    \caption{Vuong test results on dependent variables – Central African Subsample }
    \label{tab3}
\end{table}
Tables ~\ref{tab1}, ~\ref{tab2}, and~\ref{tab3} presents the Vuong test results, by considering the three predefined samples. From  table~\ref{tab1}, the probit model is only preferred for the variable $borrowed\_fin$. The logit model is then considered for the remaining variables. Similarly,  table~\ref{tab2} shows that the probit model outperforms the logit model in one occasion :  $saved$. Therefore, we consider the logit model for the remaining variables. Finally, table~\ref{tab3} shows that the probit model is preferred only for the variable $merchantpay\_dig$. As a result, we consider the logit model for the rest of the variables.

\subsection{\textit{Logit and probit models}}

Following Vuong test results, both probit and logit models are used depending on the dependent variable. Both models estimate the probability of a binary event. The logit model is represented as follows : 
\begin{align*}
    P(Y_i=1|X_i)=\frac{1}{1+e^{-X_i\beta}},
\end{align*}
similarly, the probit model is given by : 
\begin{align*}
    P(Y_i=1|X_i)=\phi(X_i\beta),
\end{align*}
where : $Y_i$ is a binary variable (e.g., owning a formal account, saving, borrowing, etc.), $X_i$ is a vector of independent variables for individual i (e.g., gender, age, age², income level, educational level, etc. ), $\beta $ is the vector of parameters to estimate, and $\phi$ is the standard normal cumulative distribution function.

\subsection{\textit{K-fold cross-validation}}

To ensure the robustness and generalizability of the binary models, we use the K-fold cross-validation approach. In this method, the dataset is randomly divided into K equal subsets (folds). For each fold, it is used once as a test set, while the remaining K–1 folds serve as the training set. The models are trained on the training data and evaluated on the test set \citep{refaeilzadeh2009cross,tajmouati2024applying}. \cite{yeh2009comparisons} in their seminal study on credit scoring models, empirically demonstrated that K-fold cross-validation significantly improves model stability and helps prevent overfitting compared to traditional validation methods. In our study, we apply a 10-fold cross-validation approach to assess the performance of each model across the full sample and the subsamples. A logit or probit model is chosen based on the outcome of the Vuong test applied to the training set and is then evaluated on the corresponding test fold. After completing all folds, the performance scores from each test set are averaged to obtain the final results. 

\begin{table}[h]
    \centering
    \begin{tabular}{c|c|c|c}
        Dependent variable & Full sample & West African subsample & Central African subsample \\ \hline
        $account\_fin$ & 0.770     & 0.761 & 0.793 \\
        $account\_mob$ & 0.738    &0.745 & 0.716 \\
        $saved\_fin$ & 0.883    & 0.873 & 0.866 \\
        $borrowed\_fin$ & 0.861   & 0.860 & 0.856 \\
        $saved$ & 0.645    & 0.643 & 0.651 \\
         $borrowed$ & 0.577    & 0.560 & 0.607  \\
          $merchantpay\_dig$ & 0.865& 0.869 & 0.847 \\
          $anydigpayment$ & 0.708    & 0.711& 0.707 \\ \hline
    \end{tabular}
    \caption{10-fold cross validation accuracy }
    \label{tab4}
\end{table}
Table~\ref{tab4} represents model accuracy across the full sample, West and Central African samples for the eight financial inclusion indicators. The accuracy is measured by the proportion of total correct predictions made by the model. Overall, the models demonstrate good predictive performance, with accuracy values mostly above 0.7.

\section{Results and discussion}
\label{sec.4}

In this section, we analyze the determinants of financial and digital inclusion across the three samples.

\subsection{\textit{Full sample}}
Tables ~\ref{tab5} and ~\ref{tab6} show the regression results across the eight financial inclusion indicators. The results show that socioeconomic, geographic, and country-level factors significantly influence financial behaviors. There is a persistent gender gap: women are 22.6\% less likely to own a financial account, 24.9\% less likely to have a mobile money account, 22.3\% less likely to save formally, and 25.3\% less likely to make digital payments. However, gender does not significantly affect inclusive saving and formal borrowing, and only modestly reduces inclusive borrowing by 10.2\%. Age has a non-linear relationship with all indicators, typically peaking in midlife before declining. Employment status plays a critical role—being unemployed reduces the odds of owning an account by 49.6\%, having a mobile money account by 45.9\%, saving formally by 56.6\%, borrowing formally by 28.2\%, saving inclusively by 55.2\%, borrowing inclusively by 38.4\%, and using digital payments by around 50–61\%. Rural residents consistently face lower odds—27.1\% less likely to own a financial account, 29.9\% less likely to have a mobile money account, 24.5\% less likely to save formally, 23.6\% less likely to save inclusively, 36.2\% less likely to make digital payments, and 17.2\% less likely to make merchant digital payments. However, rural residence does not affect significantly borrowing behaviours. Education and income are powerful predictors: individuals with only primary education are 69.8\%–89.5\% less likely to have an account, save, or use digital services than those with tertiary education. On the other hand, as compared to the tertiary education, individuals with primary education are 19.3\%-43.6\% less likely to borrow.  The lowest income group shows 10\%–66.4\% lower odds across all indicators, compared to the highest income group. Differences between countries are noticeable. Compared to Ghana, respondents from countries like Niger, and The Gambia often show drastically lower odds of financial inclusion—up to 97.8\% less likely to own mobile accounts, save, or make digital payments—while Nigerians are twice as likely to own financial accounts but 94\% less likely to use mobile money. Some countries (e.g., Liberia and Mali) show higher odds of borrowing, possibly due to stronger informal or semi-formal credit markets. In sum, women, rural residents, the uneducated, and low-income groups demonstrate the weakest financial and digital inclusion, with the lowest adoption seen in countries like the Niger.

\begin{table}[]
    \centering
    \resizebox{0.75\textwidth}{!}{%  
    \begin{tabular}{c|c|c|c|c}
    & $account\_fin$& $account\_mob$ & $saved\_fin$ & $borrowed\_fin$ \\\hline
      $female$   & $-2.567e-01^{***}$ & $-2.862e-01^{***}$ & $-0.2524009^{***}$&   $-1.754e-03$\\
      & (6.17e-11) & (7.58e-14)  & (2.29e-06)& (0.955047)\\
$age$   & $5.767e-02^{***}$ & $4.560e-02^{***}$ & $0.0407715^{***}$& $2.471e-02^{***}$  \\
      &  (< 2e-16)&  (1.17e-11)&(7.62e-06) & (2.29e-06)\\
      $age.2$   & $-4.618e-04^{***}$ & $-6.158e-04^{***}$ & $-0.0003190^{**}$&  $-2.403e-04^{***}$ \\
      & (2.19e-09) & (1.81e-13) & (0.003369)& (0.000145)\\
      $emplout$   & $-6.853e-01^{***}$ &  $-6.146e-01^{***}$  & $-0.8347075^{***}$&  $-3.309e-01^{***}$ \\
      & (< 2e-16) & (< 2e-16) & (< 2e-16)& (< 2e-16)\\
      $rural$   &  $-3.162e-01^{***}$& $-3.565e-01^{***}$ & $-0.2814556^{***}$&  $3.388e-02$ \\
      & (8.42e-14) & (< 2e-16)  &(1.52e-06) & (0.312674)\\
      $educ\_1$   & $-2.150^{***}$ &$-1.517^{***}$  & $-1.9183882^{***}$&  $-5.719e-01^{***}$ \\
      & (< 2e-16) & (< 2e-16)  &(< 2e-16)  & (< 2e-16)\\
      $educ\_2$   &  $-1.310^{***}$&  $-7.095e-01^{***}$&$-1.1128581^{***}$ & $-4.597e-01^{***}$  \\
      & (< 2e-16) & (2.80e-14) & (< 2e-16) & (4.11e-13)\\
      $inc\_1$   & $-7.161e-01^{***}$ & $-8.146e-01^{***}$ &$-1.0924808^{***}$ & $-2.115e-01 ^{***}$  \\
      & (< 2e-16) & (< 2e-16) & (< 2e-16) & (3.50e-05)\\
      $inc\_2$   & $-6.624e-01^{***}$& $-5.081e-01^{***}$ & $-0.8251729^{***}$&  $-7.197e-02$ \\
      & (< 2e-16) & (< 2e-16) & (< 2e-16) & (0.131171)\\
      $inc\_3$   & $-5.902e-01^{***}$ & $-4.125e-01^{***}$ & $ -0.8159352^{***}$&   $-9.940e-02^{*}$\\
      & (< 2e-16) & (2.49e-13) & (< 2e-16) & (0.030519)\\
      $inc\_4$   & $-4.611e-01^{***}$ & $-1.936e-01^{***}$ & $ -0.5318332^{***}$&  $-7.875e-02^{.}$ \\
      &(< 2e-16)  & (0.000241) & (1.21e-14)& (0.068483)\\
      $economycode\_BEN$   & $-6.150e-01^{***}$ & $-9.518e-01^{***}$ & $ -1.3991455^{***}$& $1.459e-01^{.}$  \\
      & (3.72e-09) & (< 2e-16) & (< 2e-16)& (0.081140 )\\
      $economycode\_BFA$   &$ -3.844e-01^{***}$ & $-1.282^{***}$ & $ -0.7000616^{***}$& $1.558e-02$  \\
      &  (0.000185)& (< 2e-16) & (1.47e-07)& (0.859059)\\
      $economycode\_CIV$   & $-5.975e-01^{***}$ &  $-7.015e-01^{***}$& $ -0.9895592^{***}$&  $ -2.341e-01^{*}$ \\
      & (8.20e-09) & (2.42e-12) &(7.10e-13) & (0.013251) \\
      $economycode\_CMR$   & $-6.062e-01^{***}$ & $-8.000e-01^{***}$ &  $ -0.9319117^{***}$&  $7.548e-02$ \\
      & (3.65e-09) & (1.23e-15) & (3.66e-12)& (0.377210) \\
      $economycode\_COD$   &  $-1.950^{***}$&  $-1.837^{***}$ & $-1.8510606^{***}$& $-2.576e-01^{**}$  \\
      & (< 2e-16) &  (< 2e-16)& (< 2e-16)& (0.005562)\\
      $economycode\_COG$   & $-1.055^{***}$ & $-1.837e+00^{***}$ & $-1.5953684^{***}$& $-1.343e-01$   \\
      & (< 2e-16) &  (< 2e-16)&(< 2e-16) & (0.143510)\\
      $economycode\_GIN$   & $-1.404^{***}$ &  $-1.571^{***}$  & $-1.5447275^{***}$& $-3.900e-02 $ \\
      & (< 2e-16) &  (< 2e-16)& (< 2e-16)& (0.660786)\\
      $economycode\_GMB$   & $-1.897e-01^{.}$ & $-3.838^{***}$ & $-0.1989306^{.}$& $-3.462e-02$  \\
      & (0.059203) & (< 2e-16) & (0.096543)& (0.698479)\\
      $economycode\_LBR$   & $-3.869e-01^{***}$& $-8.031e-01^{***}$ & $-0.5698752^{***}$&  $4.864e-01^{***}$ \\
      & (0.000196) & (2.00e-15) & (1.42e-05)& (8.29e-10)\\
      $economycode\_MLI$   & $-2.117e-01^{*}$ & $-1.120^{***}$ & $-0.3897181^{**}$&  $2.450e-01^{**}$ \\
      & (0.040050) &  (< 2e-16)& (0.002384)& (0.003166)\\
      $economycode\_NER$   &  $-1.421^{***}$&  $-3.083^{***}$& $-1.5846023^{***}$&  $-2.343e-01^{*}$ \\
      & (< 2e-16) & (< 2e-16) & (< 2e-16)& (0.019359)\\
      $economycode\_NGA$   & $7.127e-01^{***}$&  $-2.818^{***}$& $0.1724641$& $-5.399e-02$  \\
      & (5.77e-13) & (< 2e-16) & (0.113826)& (0.536680)\\
      $economycode\_SEN$   &  $-2.503e-01^{*}$&  $-3.046e-01^{**}$&$-0.6619464^{***}$ &  $1.642e-01^{.}$ \\
      & (0.013378) &  (0.002536)& (2.64e-07)& (0.050781) \\
      $economycode\_SLE$   & $-1.145^{***}$ & $-1.800^{***}$ &$-1.2403172^{***}$ &  $-2.358e-01^{*}$ \\
      & (< 2e-16) &  (< 2e-16)& (3.48e-16)& (0.012774)\\
      $economycode\_TCD$   & $-6.014e-01^{***}$& $-2.180^{***}$ & $-0.7294790^{***}$& $2.580e-01^{**}$  \\
      & (1.99e-08) & (< 2e-16) & (1.47e-07)& (0.002073)\\
      $economycode\_TGO$   &  $-3.627e-01^{***}$&  $-9.473e-01^{***}$& $-0.4787808^{***}$&  $3.592e-02$ \\
      & (0.000382) & (< 2e-16) & (1.47e-07)&(0.675541) \\
      \end{tabular}%
      }
    \caption{Regression analysis – full sample}
    \label{tab5}
\end{table}

\begin{table}[]
    \centering
    \resizebox{0.75\textwidth}{!}{%  
    \begin{tabular}{c|c|c|c|c}
    & $borrowed$& $saved$ & $anydigpayment$ & $merchantpay\_dig$  \\\hline
      $female$   & $-1.080e-01^{***}$ & $6.819e-03$ & $-2.919e-01^{***}$&  $-0.2949249^{***}$ \\
      & (0.000931)  & (0.84220) & (1.28e-15)&(6.26e-07) \\
$age$   & $4.784e-02^{***}$ & $ 2.967e-02^{***}$ &$5.518e-02 ^{***} $& $ 0.0097266$ \\
      & (< 2e-16) & (5.23e-08) & (< 2e-16)& (0.386062)\\
      $age.2$   & $-6.103e-04^{***}$ &$-3.771e-04^{***}$  &$-5.993e-04^{***}$ & $-0.0002580^{.}$  \\
      & (< 2e-16) & (1.09e-08) & (6.82e-16 )& (0.072432)\\
      $emplout$   & $4.850e-01^{***}$ & $-8.033e-01^{***}$ &-$6.968e-01^{***}$ & $-0.9349250^{***}$  \\
      &  (< 2e-16)& (< 2e-16) & (< 2e-16)& (< 2e-16)\\
      $rural$   & $4.312e-02$ &  $-2.691e-01^{***}$& $-4.493e-01^{***}$&  $-0.1888073^{**}$ \\
      & (0.219491) & (2.18e-13) &(< 2e-16) &(0.003195) \\
      $educ\_1$   & $-2.149e-01^{*}$ & $-1.199^{***}$ & $-2.253^{***}$&  $-1.6422720^{***}$ \\
      & (0.012693) & (< 2e-16) &(< 2e-16) &(< 2e-16)\\
      $educ\_2$   & $-1.432e-01^{.}$ & $-6.529e-01^{***}$ &$-1.283^{***}$ &  $-0.9697620^{***}$ \\
      & (0.085312) & (7.30e-11) & (< 2e-16)& (< 2e-16)\\
      $inc\_1$   & $-1.050e-01^{*}$ & $-8.579e-01^{***}$ &$-9.470e-01^{***}$ &  $-0.6286177^{***}$ \\
      & (0.040241) & (< 2e-16) &(< 2e-16) &(4.21e-10) \\
      $inc\_2$   & $1.338e-01^{**}$ & $-5.091e-01^{***}$ & $-6.616e-01^{***}$&  $-0.4761298^{***}$ \\
      & (0.008752) &  (< 2e-16)& (< 2e-16)& (3.25e-07 ) \\
      $inc\_3$   & $7.391e-02$ & $-3.698e-01^{***}$ & $-5.478e-01^{***}$&  $-0.4948553^{***}$ \\
      &  (0.130327)& (4.99e-13) & (< 2e-16)& (1.49e-08)\\
      $inc\_4$   & $7.572e-02 $ & $-2.447e-01^{***}$ & $-3.531e-01^{***}$&  $-0.3184566^{***}$ \\
      & (0.104561) & (6.84e-07)  & ( 6.50e-12)& (3.95e-05)\\
      $economycode\_BEN$   & $-1.538e-01^{.}$ & $-8.291e-01 ^{***}$ &$-8.432e-01^{***}$ & $-1.2537199^{***}$  \\
      & (0.092614) &(< 2e-16)  & (6.73e-16 )& (3.85e-14)\\
      $economycode\_BFA$   & $-8.257e-02$ & $ -7.053e-01^{***}$ &$-1.019^{***}$ &  $-0.8814430^{***}$ \\
      & (0.369587) &  (2.80e-12)& (< 2e-16)& (1.36e-08 )\\
      $economycode\_CIV$   & $-2.214e-01^{*}$ & $-9.083e-01^{***}$ &$-5.963e-01^{***}$ & $-0.6143906^{***}$  \\
      & (0.015767) & (< 2e-16) & (1.17e-08)& (1.42e-05)\\
      $economycode\_CMR$   & $3.275e-01^{***}$ & $-4.238e-01^{***}$ &$-7.254e-01^{***}$ &  $-0.4732770^{***}$ \\
      &(0.000396)  & (3.20e-05) & (4.08e-12)& (0.000431)\\
      $economycode\_COD$   & $8.231e-02$ & $-1.351^{***}$ & $-2.028^{***}$&  $-1.4011290^{***}$ \\
      & (0.370262) & (< 2e-16) & (< 2e-16 )& (< 2e-16)\\
      $economycode\_COG$   & $-3.140e-01^{***}$ &  $-9.569e-01^{***}$&$-1.026^{***}$ &  $-0.8058289^{***}$ \\
      & (0.000689) &  (< 2e-16)& (< 2e-16 )& (4.99e-08)\\
      $economycode\_GIN$   & $2.982e-01^{**}$ &  $-6.947e-01^{***}$&$-1.505^{***}$ &  $-0.7626672^{***}$ \\
      & (0.001344) & (8.24e-12) & (< 2e-16 )& (2.63e-07)\\
      $economycode\_GMB$   & $1.835e-01^{*}$ & $-7.603e-01^{***}$ &$ -2.153^{***}$&   $-1.5064193^{***}$\\
      & (0.046469) & (4.86e-14) & (< 2e-16 )& (5.95e-16)\\
      $economycode\_LBR$   & $4.524e-01^{***}$ & $-2.426e-01^{*}$  &$-6.143e-01^{***}$ &  $-0.8322394^{***}$ \\
      & (1.47e-06) &  (0.01795)& (4.93e-09)& (7.47e-08)\\
      $economycode\_MLI$   & $-1.374e-01$ & $-3.299e-01^{**}$ & $-9.239e-01^{***}$&  $-0.1258209$ \\
      & (0.137851) & (0.00125) & (< 2e-16)& (0.346473)\\
      $economycode\_NER$   & $3.495e-01^{***}$ & $-1.580^{***}$ &$-2.669^{***}$ &   $-1.4538578^{***}$\\
      &(0.000206)  & (< 2e-16) & (< 2e-16)& (1.27e-12)\\
      $economycode\_NGA$   & $1.166e-01$ & $-5.021e-01^{***}$ & $-1.181^{***}$& $-0.1399241$  \\
      &(0.204558)  & (9.99e-07) &(< 2e-16) & (0.246950)\\
      $economycode\_SEN$   & $2.544e-02$ & $-2.953e-01^{**}$ & $-2.094e-01^{*}$&  $0.0691994$ \\
      & (0.782088)& (0.00403) & (0.0492)& (0.578957)\\
      $economycode\_SLE$   & $1.036e-01$ & $-6.568e-01^{***}$ & $-1.546^{***}$&   $-1.4132849^{***}$\\
      & (0.260277) & (6.89e-11) & (< 2e-16)& (2.15e-14)\\
      $economycode\_TCD$   & $1.454e-01 $ & $-1.084^{***}$ &$-1.880^{***}$ &   $-0.4377826^{**}$\\
      & (0.117558) & (< 2e-16) & (< 2e-16)& (0.002051)\\
      $economycode\_TGO$   & $-2.111e-01^{*}$ & $ -5.435e-01^ {***}$ &$ -8.503e-01^{***}$&   $-1.5871782^{***}$\\
      & (0.021564) & (9.43e-08) & (6.59e-16)& (< 2e-16)\\
      \end{tabular}%
      }
    \caption{Regression analysis – full sample}
    \label{tab6}
\end{table}

\begin{table}[]
    \centering
    \resizebox{0.75\textwidth}{!}{%  
    \begin{tabular}{c|c|c|c|c}
    & $account\_fin$& $account\_mob$ & $saved\_fin$ & $borrowed\_fin$ \\\hline
      $female$   & $-2.893e-01^{***}$ & $-3.402e-01^{***}$ &$-0.3178793^{***}$ & $0.0474779$  \\
      & (5.94e-11) & (1.37e-14) & (7.97e-08)&(0.507097) \\
$age$   & $6.122e-02^{***}$ &  $4.555e-02^{***}$&$0.0409374^{***}$ & $0.0541057 ^{***}$  \\
      & (< 2e-16) &  (2.40e-09) & (3.67e-05)&(9.84e-06) \\
      $age.2$   & $-5.114e-04^{***}$ & $-5.767e-04^{***}$ & $-0.0002834^{*}$&  $-0.0005040^{***}$ \\
      & (4.55e-09) &  (9.78e-10) & (0.015610)&(0.000600) \\
      $emplout$   & $-6.528e-01^{***}$ & $-5.547e-01^{***}$ & $-0.7368327^{***}$&  $-0.7249052 ^{***}$ \\
      & (< 2e-16) & (< 2e-16) & (< 2e-16)&(1.33e-12)\\
      $rural$   & $-2.941e-01^{***}$ & $-3.527e-01^{***}$ & $-0.2850864^{***}$&   $0.1583004^{*}$\\
      & (6.27e-10) &  (6.11e-14) & (1.04e-05)&(0.041008) \\
      $educ\_1$   & $-2.296^{***}$ & $-1.377^{***}$ &$-1.9639297^{***}$ &$-1.1899746 ^{***}$   \\
      & (< 2e-16) & (< 2e-16) &(< 2e-16) &(< 2e-16) \\
      $educ\_2$   & $-1.463^{***}$ & $-5.798e-01^{***}$ &$-1.0372042^{***}$ &$-0.9422920^{***}$   \\
      & (< 2e-16) &  (2.49e-07) & (< 2e-16)&(7.09e-13)\\
      $inc\_1$   & $-7.068e-01^{***}$ &  $-8.615e-01^{***}$& $-1.0945092^{***}$& $-0.4449588^{***}$  \\
      & (< 2e-16) &  (< 2e-16) & (< 2e-16)&(0.000215)\\
      $inc\_2$   & $-6.948e-01^{***}$ &  $-4.492e-01^{***}$&$-0.8195483^{***}$ & $-0.1584589$  \\
      &  (< 2e-16)&  (5.06e-11) & (< 2e-16)&(0.146124 )\\
      $inc\_3$   & $-6.010e-01^{***}$ & $-4.250e-01^{***}$ &$-0.8131789^{***}$ &$-0.2336279^{*}$   \\
      & (< 2e-16) &  (6.70e-11) & (< 2e-16)&(0.027516) \\
      $inc\_4$   & $-4.672e-01^{***}$ & $-2.295e-01^{***}$ & $-0.5016701^{***}$&$-0.1642265^{.}$   \\
      & (1.54e-14) &  (0.000186) & (5.41e-11)&(0.096708)\\
      $economycode\_BEN$   & $-3.713e-01^{***}$ & $-6.310e-01^{***}$ &$-0.7321746^{***}$ & $-0.0511447$  \\
      &(0.000525)  &  (1.06e-10) & (9.35e-06)&(0.750443) \\
      $economycode\_BFA$   & $-1.327e-01$ & $-9.685e-01^{***}$ &$-0.0258860$ &$-0.3220084^{.}$   \\
      &(0.206079)  &  (< 2e-16) & (0.860279)&(0.062647) \\
      $economycode\_CIV$   & $-3.517e-01^{***}$ &$-3.920e-01^{***}$  &$-0.3294528^{*}$ &$-0.8278680^{***}$   \\
      & (0.000914) & (5.34e-05) &(0.029684) &(2.01e-05 ) \\
     $economycode\_GHA$   & $2.511e-01^{*}$ &$3.215e-01^{**}$  &$0.6555361^{***}$ &$-0.3287821 ^{*}$   \\
      & (0.013332) & (0.001424) & (3.74e-07)&(0.049996) \\
      
      $economycode\_GIN$   & $-1.187e+00^{***}$ &$-1.244^{***}$  &$-0.8713182^{***}$ &$-0.4429859^{*}$   \\
      &  (< 2e-16)& (< 2e-16) & (6.45e-07)&(0.012130) \\
      $economycode\_GMB$   & $6.516e-02$ &$-3.516^{***}$  &$0.4653208^{***}$ &$-0.4061451^{*}$   \\
      & (0.525697) &  (< 2e-16) & (0.000556)&(0.022362) \\
      $economycode\_LBR$   & $-1.398e-01$ &$-4.865e-01^{***}$  &$0.0986499$ &$0.5690353^{***}$   \\
      & (0.190058) &  (7.88e-07) & (0.500204)&(0.000117) \\
      $economycode\_MLI$   & $ 3.671e-02 $ & $-8.097e-01^{***}$ &$0.2948062^{*}$ &$0.1203393$   \\
      &(0.726705)  &  (7.55e-16) & (0.038536)&(0.445030) \\
      $economycode\_NER$   & $-1.168^{***}$ &$-2.777^{***}$  &$-0.9064262^{***}$ &   $-0.8282718^{***}$\\
      & (< 2e-16) & ( < 2e-16) & (2.91e-06)&(7.91e-05) \\
      $economycode\_NGA$   & $9.654e-01^{***}$ &$-2.489^{***}$  &$ 0.8170472^{***}$ &$-0.4607225^{**}$   \\
      &  (< 2e-16)&  ( < 2e-16) & (1.33e-10)&(0.008388)\\
      
      $economycode\_SLE$   & $-8.977e-01^{***}$ &$-1.482^{***}$  &$-0.5689877^{***}$ & $-0.8995725^{***}$  \\
      & (1.10e-14) &  ( < 2e-16) & (0.000555)&(5.36e-06)\\
      
      $economycode\_TGO$   & $-1.171e-01$ &$-6.287e-01 ^{***}$  &$0.1849064$ &   $-0.2741959$\\
      & (0.266021 ) &  (1.97e-10 ) & (0.184561)&(0.101671) \\
      \end{tabular}%
      }
    \caption{Regression analysis –  West African subsample}
    \label{tab7}
\end{table}

\begin{table}[]
    \centering
    \resizebox{0.75\textwidth}{!}{%  
    \begin{tabular}{c|c|c|c|c}
    & $borrowed$& $saved$ & $anydigpayment$ & $merchantpay\_dig$  \\\hline
      $female$   & $-8.920e-02^{*}$ & $-5.959e-03$ &$-3.724e-01^{***}$ & $-0.3961635^{***}$  \\
      & (0.016563) & (0.80326) & (< 2e-16)&(6.46e-09) \\
$age$   & $4.484e-02^{***}$ &  $1.866e-02^{***}$&$5.785e-02^{***}$ & $-0.0014310$  \\
      & (6.00e-14) &  (6.38e-07) & (< 2e-16)&(0.9084) \\
      $age.2$   & $-5.709e-04 ^{***}$ & $-2.341e-04^{***}$ & $ -5.999e-04 ^{***}$&  $-0.0001015$ \\
      & (3.08e-15) &  (2.11e-07) & (8.28e-13)&(0.5166) \\
      $emplout$   & $-4.291e-01 ^{***}$ & $-4.683e-01^{***}$ & $ -6.219e-01^{***}$&  $-0.7974895^{***}$ \\
      & (< 2e-16) & (< 2e-16) & (< 2e-16)&(2.76e-16)\\
      $rural$   & $1.307e-01^{**}$ & $-1.382e-01^{***}$ & $-4.445e-01^{***}$&   $-0.1631816^{*}$\\
      & (0.001088) &  (6.79e-08) & (< 2e-16)&( 0.0257) \\
      $educ\_1$   & $-2.348e-01^{*}$ & $-6.673e-01^{***}$ &$-2.168^{***}$ &$-1.7160341^{***}$   \\
      & (0.019266) & (< 2e-16) &(< 2e-16) &(< 2e-16) \\
      $educ\_2$   & $-1.368e-01$ & $-3.102e-01^{***}$ &$-1.204^{***}$ &$-1.0039514^{***}$   \\
      & (0.164178) &  (7.57e-06) & (< 2e-16)&(< 2e-16)\\
      $inc\_1$   & $-1.105e-01^{.}$ &  $-5.155e-01^{***}$& $-9.706e-01^{***}$& $-0.6310160^{***}$  \\
      & (0.058162) &  (< 2e-16) & (< 2e-16)&(3.87e-08)\\
      $inc\_2$   & $1.080e-01^{.}$ &  $-2.792e-01^{***}$&$-6.144e-01^{***}$ & $-0.5460563^{***}$  \\
      &  (0.063184)&  (5.23e-14) & (< 2e-16)&(5.00e-07)\\
      $inc\_3$   & $5.201e-02^{}$ & $-2.109e-01^{***}$ &$-5.720e-01^{***}$ &$-0.4895136 ^{***}$   \\
      & (0.350076) &  (3.66e-09) & (< 2e-16)&(8.63e-07) \\
      $inc\_4$   & $7.827e-03$ & $-1.670e-01^{***}$ & $-3.919e-01 ^{***}$&$-0.3662598^{***}$   \\
      & (0.883475) &  (1.30e-06) & (4.05e-11)&(4.51e-05)\\
      $economycode\_BEN$   & $-1.876e-01^{*}$ & $-3.217e-01^{***}$ &$-6.199e-01^{***}$ & $-1.3297051^{***}$  \\
      &(0.039685)  &  (4.42e-08) & (6.50e-10)&(5.05e-15) \\
      $economycode\_BFA$   & $-1.206e-01$ & $-2.488e-01^{***}$ &$-8.000e-01^{***}$ &$-0.9492852^{***}$   \\
      &(0.187358)  &  (2.39e-05) & (2.05e-15)&(2.04e-09) \\
      $economycode\_CIV$   & $-2.563e-01^{**}$ &$-3.737e-01^{***}$  &$-3.828e-01 ^{***}$ &$-0.6851777^{***}$   \\
      & (0.004996) & (2.07e-10) &(0.000134) & (2.41e-06)\\
     $economycode\_GHA$   & $-2.139e-02$ &$1.806e-01^{**}$  &$2.250e-01  ^{*}$ &$-0.0671432$  \\
      & (0.816072) & (0.00327) & ( 0.034654 )&(0.5915)  \\
      
      $economycode\_GIN$   & $2.593e-01^{**}$ &$-2.414e-01^{***}$  &$-1.286^{***}$ &$-0.8481223^{***}$   \\
      &  (0.005092)& ( 4.65e-05) &(< 2e-16) & (2.92e-08)\\
      $economycode\_GMB$   & $1.651e-01^{.}$ &$-2.772e-01^{***}$  &$-1.934^{***}$ &$-1.5714512^{***}$   \\
      & (0.071456) &  (2.43e-06) & (< 2e-16)&(< 2e-16) \\
      $economycode\_LBR$   & $4.050e-01^{***}$ &$3.019e-02$  &$-3.917e-01^{***}$ &$-0.9011267^{***}$   \\
      & ( 1.52e-05) &  (0.61343) & (0.000106)&(1.40e-08) \\
      $economycode\_MLI$   & $ -1.759e-01^{.} $ & $-1.855e-02$ &$-7.099e-01^{***}$ &$-0.1862932$   \\
      &(0.055710)  &  (0.75500) & (2.53e-12)&(0.1731) \\
      $economycode\_NER$   & $3.269e-01^{***}$ &$-7.800e-01^{***}$  &$-2.462^{***}$ &   $-1.5131861^{***}$\\
      & (0.000421) & (< 2e-16) & (< 2e-16)&(2.09e-13) \\
      $economycode\_NGA$   & $9.354e-02$ &$ -1.254e-01^{*}$  &$-9.568e-01^{***}$ &$-0.2100046 ^{.}$   \\
      &  (0.313132 )&  (0.03739) & (< 2e-16)&(0.0987)\\
      
      $economycode\_SLE$   & $6.407e-02$ &$-2.197e-01^{***}$  &$-1.324^{***}$ & $-1.4737643^{***}$  \\
      & (0.484912) &  ( 0.00019) & (< 2e-16)&(4.85e-15)\\
      
      $economycode\_TGO$   & $-2.541e-01^{**}$ &$-1.528e-01^{*}$  &$-6.295e-01 ^{***}$ &   $-1.6673132^{***}$\\
      & (0.005789 ) &  (0.01039 ) & (6.16e-10)&(< 2e-16 ) \\
      \end{tabular}%
      }
    \caption{Regression analysis –  West African subsample}
    \label{tab8}
\end{table}

\begin{table}[]
    \centering
    \resizebox{0.75\textwidth}{!}{%  
    \begin{tabular}{c|c|c|c|c}
    & $account\_fin$& $account\_mob$ & $saved\_fin$ & $borrowed\_fin$ \\\hline
      $female$   & $-0.1407708^{.}$ & $-0.1293410^{.}$ &$0.0352933$ & $-0.2148636$  \\
      & (0.0974) & (0.093609) & (0.77719 )&(0.105446) \\
$age$   & $0.0458497^{**}$ &  $0.0488519^{***}$ & $0.0668995^{**}$ & $-0.0393172^{***}$  \\
      & (1.01e-08) &  (0.000674) & (0.00601)&(0.000497) \\
      $age.2$   & $-0.0002936 ^{***}$ & $-0.0007821^{***}$ & $-0.0008309^{**}$&  $0.0004352^{**}$ \\
      & (0.0055 ) &  (1.84e-05) & (0.00725)&(0.005484) \\
      $emplout$   & $-0.7982447^{***}$ & $-0.7704886^{***}$ & $-1.2621308^{***}$ &  $-0.7172360^{***}$ \\
      & (5.42e-15) & (< 2e-16) & (1.51e-11)&(1.81e-05)\\
      $rural$   & $-0.3984741^{***}$ & $-0.3497947^{***}$ &$-0.2377521^{.}$&   $-0.1967158$\\
      & (2.27e-05) &  (3.28e-05) & ( 0.08794)&(0.175710) \\
      $educ\_1$   & $-1.8464762^{***}$ & $-1.8864053^{***}$ &$-1.5134225^{***}$ &$-1.3021600^{***}$   \\
      & (< 2e-16) & (< 2e-16) &(4.96e-10) &(2.10e-07) \\
      $educ\_2$   & $-1.0023940^{***}$ & $-1.0007059^{***}$ &$-1.3141634^{***}$ &$-1.1361930^{***}$   \\
      & (5.78e-12) &  (2.49e-09) & (1.00e-10)&(4.52e-08)\\
      $inc\_1$   & $-0.7767760^{***}$ &  $-0.6851572^{***}$& $-1.1314109^{***}$& $-0.5146443^{*}$  \\
      & (3.79e-08) &  (4.49e-08) & (4.78e-07)&(0.023859)\\
      $inc\_2$   & $-0.5485099^{***}$ &  $-0.7228991^{***}$&$ -0.8704494^{***}$ & $-0.2046485$  \\
      &  ( 3.40e-05)&  (4.15e-09) & (1.56e-05)&(0.313506)\\
      $inc\_3$   & $-0.5595410^{***}$ & $-0.3772077^{***}$ &$-0.8477259 ^{***}$ &$-0.0902562$   \\
      & (7.57e-06) &  (0.000851) & (5.73e-06)&(0.633759) \\
      $inc\_4$   & $-0.4567495^{***}$ & $-0.1132546$ & $-0.7053518 ^{***}$&$ -0.2130668$   \\
      & (5.15e-05) &  (0.274630) & (1.46e-05 )&(0.236421)\\
      $economycode\_COD$   & $-1.2619270^{***}$ & $-1.0951093^{***}$ &$-0.8406659^{***}$ & $-0.9675356^{***}$  \\
      &(< 2e-16)  &  (< 2e-16) & (4.23e-06)&(5.13e-07) \\
      $economycode\_COG$   & $-0.5590128^{***}$ & $-0.2115095 ^{*}$ &$-0.5335302^{**}$ &$-0.5203523^{**}$   \\
      &(8.48e-07)  &  (0.034504) & (0.00229)&(0.005794) \\
      $economycode\_TCD$   & $0.0213156$ &$-1.3688960^{***}$  &$0.0622884$ &$0.2365834$   \\
      & (0.8506) & (< 2e-16) & (0.70134) & (0.150209) \\
    
      \end{tabular}%
      }
    \caption{Regression analysis –  Central African subsample}
    \label{tab9}
\end{table}

\begin{table}[]
    \centering
    \resizebox{0.75\textwidth}{!}{%  
    \begin{tabular}{c|c|c|c|c}
    & $borrowed$& $saved$ & $anydigpayment$ & $merchantpay\_dig$  \\\hline
      $female$   & $-0.1852562^{**}$ & $0.0450281$ &$-0.0359410$ & $0.0148250$  \\
      & (0.00656) & (0.5255 ) & (0.63495 )&(0.810721) \\
$age$   & $0.0619912^{***}$ &  $0.0276463  ^{*}$&$ 0.0504180^{***}$ & $0.032294^{*}$  \\
      & (9.50e-08) &  (0.0179) & (0.00012)&(0.012046) \\
      $age.2$   & $-0.0007896^{***}$ & $-0.0003653^{*}$ & $-0.0006478^{***}$&  $-0.0005114^{**}$ \\
      & (3.39e-08) &  (0.0104) & (6.33e-05)&(0.002738) \\
      $emplout$   & $-0.6464520^{***}$ & $-0.9205871^{***}$ & $-0.9084588^{***}$&  $-0.6703899^{***}$ \\
      & (< 2e-16) & (< 2e-16) & (< 2e-16)&(1.83e-15)\\
      $rural$   & $-0.2483570^{***}$ & $-0.4136001^{***}$ & $-0.4484425^{***}$&   $-0.1402538^{*}$\\
      & (0.00078) &  (5.64e-08) & (3.84e-08)&(0.039578) \\
      $educ\_1$   & $-0.0976876$ & $-1.3293520^{***}$ &$-2.4584916^{***}$ &$-0.7162544^{***}$   \\
      & (0.57097) & (5.39e-12) &(< 2e-16) &( 1.46e-07) \\
      $educ\_2$   & $-0.1256740$ & $-0.8971542^{***}$ &$ -1.4423750^{***}$ &$-0.4537015 ^{***}$   \\
      & (0.42564) &  (5.52e-07) & (5.08e-14)&(0.000124)\\
      $inc\_1$   & $-0.1133678$ &  $-0.9418038^{***}$& $-0.8908955^{***}$& $-0.3434310^{***}$  \\
      & (0.29135) &  (< 2e-16) & (2.11e-13)&(0.000965)\\
      $inc\_2$   & $0.2056928^{.}$ &  $-0.7059694 ^{***}$&$-0.8523921^{***}$ & $-0.1442993$  \\
      &  (0.05431)&  (1.56e-10) & (7.35e-13)&(0.125337)\\
      $inc\_3$   & $0.1386268$ & $-0.4520461^{***}$ &$-0.4739585 ^{***}$ &$-0.2841706^{**}$   \\
      & (0.17646) &  (1.79e-05) & (1.98e-05)&(0.002437) \\
      $inc\_4$   & $0.2792723^{**}$ & $ -0.1690704^{.}$ & $-0.2631909^{*}$&$-0.1181581$   \\
      & (0.00389) &  (0.0893) & (0.01107)&(0.147568)\\
      $economycode\_COD$   & $-0.3019322^{**}$ & $-0.9883096^{***}$ &$-1.3219775^{***}$ & $-0.4362554^{***}$  \\
      &(0.00194)  &  ( < 2e-16) & (< 2e-16)&(1.27e-06 ) \\
      $economycode\_COG$   & $ -0.6516549^{***}$ & $-0.5305208^{***}$ &$-0.2460811^{*}$ &$-0.1392953^{.}$   \\
      &(6.98e-12)  &  (7.55e-08) & ( 0.01494)&(0.096602) \\
      $economycode\_TCD$   & $-0.1920686 ^{*}$ &$-0.6903277^{***}$  &$-1.1331789^{***}$ &$0.0264002$   \\
      & (0.04650) & (5.48e-12) & (< 2e-16) & (0.751621 ) \\
     
      \end{tabular}%
      }
    \caption{Regression analysis –  Central African subsample}
    \label{tab10}
\end{table}

\subsection{\textit{West African subsample} }

Tables~\ref{tab7} and ~\ref{tab8} show the regression results over the West African sample. Gender disparities are pronounced across nearly all eight indicators: women are 25.1\% less likely to own a financial account, 28.8\% less likely to have a mobile money account, 31.1\% less likely to make any digital payment, and 32.7\% less likely to make merchant digital payments. On the other hand, gender does not affect significantly formal borrowing and inclusive saving. Rural residents face major barriers—e.g., they are 29.7\% less likely to use mobile money and 35.9\% less likely to make digital payments. Age effects show a diminishing return across indicators, with digital merchant payments unaffected by age, suggesting equal convenience for all age groups. Education is a strong determinant:  Compared to individuals with tertiary education, those with only primary education are 89.9\% less likely to own a financial account, 74.7\% less likely to use mobile money, 88.6\% less likely to make digital payments, 86\% less likely to save formally, and 69.6\% less likely to borrow formally. Income plays a significant role as well— compared to the highest income group, individuals in the lowest quintile are 50.7\% less likely to own an account, 57.8\% less likely to use a mobile money account, 66.5\% are less likely to save formally, 35.9\% are less likely to borrow formally,  and 62.1\% less likely to transact digitally. Compared to Senegal, financial activity is significantly lower in Niger, Guinea, and Sierra Leone, though these countries show relatively better outcomes in borrowing—suggesting the importance of informal credit. The Gambia and Mali show slightly higher odds of account ownership (+6.7\% and +3.8\%). Nigeria has the highest rate of financial account ownership but lags in mobile money usage, similar to The Gambia, where mobile account ownership is 97\% lower than in Senegal. Overall, Senegal outperforms most West African countries in digital usage, second only to Ghana. 

\subsection{\textit{Central African subsample} }

Tables ~\ref{tab9} and  ~\ref{tab10} present the regression results for the Central African sample. Among all indicators, only inclusive borrowing shows a significant gender effect, reflecting women’s heavy reliance on informal credit. When it comes to residence area and employment, Central Africa shows more alarming patterns in financial and digital inclusion than West Africa: rural residents are up to 36.1\% less likely to access to financial and digital services, while unemployment reduces participation by 47.6\%-71.7\%, - stronger effect than observed in West Africa. Similarly, compared to the highest income quintile group, individuals in the poorest income quintile are up to 67.7\% less likely to participate financially and digitally. In almost all cases, age positively correlates with financial activity, but the diminishing effect over time is evident through the negative squared age term. Country-level differences are also significant. Compared to Cameroon, respondents in Democratic Republic of the Congo are 71.6\% less likely to have a financial account, are 66.5\% less likely to own a mobile money account, 62.8\% less likely to save inclusively, and 73.4\% to make digital payments.  The results are similar poor for Republic of the Congo and Chad. Overall, Cameroon remains the most inclusive country  in Central Africa.

\newpage
\section{Conclusion and recommendations}
\label{sec.5}

As Africa quests for more financial and digital inclusion, the West and Central regions should be one of the areas of focus, and it's confirmed by the results. We find that socio-economic and demographic variables, particularly gender, age, education, income, and rural residence, are all significant in determining both financial and digital inclusion outcomes in these two regions. However, key differences exist between sub-regions: in Central Africa, gender loses significance when it comes to formal account ownership, formal savings and borrowings, and digital payments due to high exclusion associated with access to education, income, and jobs for women and other marginalized individuals. Having a high rate of formal financial account ownership does not necessarily translate into higher digital payment usage in the two regions, whereas mobile money emerges as a more effective driver of digital inclusion. Ghana stands out as the most financially inclusive country overall in our study, but it lags behind Liberia and Senegal in credit accessibility and digital merchant payments, respectively. While Nigeria leads in formal financial account ownership and savings, it shows very low results when it comes to mobile money account ownership and digital payments compared to Ghana. The K-fold cross-validation shows the presence of extreme class imbalances in formal financial borrowing due to the extremely low access to this form of credit. This revealed the limited accessibility of formal financial credit.\\
Based on these findings and the overall comparative analysis of the three samples, it is clear that improving both financial and digital inclusion in West and Central Africa requires a combination of individual and collaborative efforts. As it will be crucial to reducing the persistent socio-economic and regional disparities across the two sub-regions. To address these gaps, policymakers should first prioritize access to free education, especially at the tertiary level, as it plays a crucial role in empowering marginalized groups. Increasing literacy rates will not only enhance individual capabilities but will also raise the likelihood of adopting and using formal financial services. Second, policymakers should consider investing in youth empowerment programs through entrepreneurship and skills training centers, with a special focus on including women and other marginalized groups. This would create future employment opportunities and a stable income source, thereby helping to close the wide gender gap seen especially in the Central African countries.\\
To promote digital financial inclusion, mobile money adoption should be further encouraged. From our results, we can see that it plays a vital role in expanding digital merchant payments and facilitating broader financial transactions. In addition to that, microfinance institutions, fintech companies, and credit unions should be supported, as they have proven more effective than mobile money alone in reaching women and other marginalized populations.\\
Access to credit remains a major problem, as many individuals in both regions avoid formal borrowing due to high collateral requirements, elevated interest rates, religious concerns, and other personal reasons. Policymakers must design more inclusive and diversified financial systems tailored to the cultural and religious contexts of local communities. For example, governments could introduce digital credit scoring systems that assess users based on mobile money transactions, utility bill payments, and other financial behaviours, offering safer, more accessible lending alternatives.\\
The results seen in Ghana show that achieving a balance between digital and formal financial inclusion is possible, so other countries should be encouraged to adopt such successful models by expanding both bank and mobile money account ownership.\\
Finally, South-South partnerships should be encouraged, in form of conferences and workshops, where countries facing similar challenges can exchange knowledge and replicate effective strategies. These efforts will contribute to broader financial inclusion across Sub-Saharan Africa, ensuring that every individual, male or female, rich or poor, living in rural or urban areas, will have equal access to financial services.
\subsection*{Disclosure Statement}
We acknowledge that there are no competing interests between the authors.

\subsection*{Funding}
This research received no external funding.

\subsection*{Data availability statement}

For our analysis, we employ the Global Findex Database 2021, available at : \\https://doi.org/10.48529/jq97-aj70

%\section*{References}
%\bibliographystyle{asa}
%\bibliographystyle{apacite}
\bibliography{references}
 \end{document}